\documentclass[12pt]{article}
\usepackage{amsmath}
\usepackage{amsfonts}

\usepackage{mltex}

\newtheorem{lemma}{Lemma}[section]
\newtheorem{theorem}[lemma]{Theorem}
\newtheorem{proposition}[lemma]{Proposition}
\newtheorem{corollary}[lemma]{Corollary}
\newtheorem{remark}[lemma]{Remark}

\newtheorem{definition}[lemma]{Definition}

\def\sq{\hbox {\rlap{$\sqcap$}$\sqcup$}}
\def\sq{\hbox {\rlap{$\sqcap$}$\sqcup$}}
\def\R{\mathbb R}
\def\C{\mathbb C}
\def\Z{\mathbb Z}
\def\N{\mathbb N}
\def\1{{\rm 1\mskip-4.5mu l} }
\def\lsim{\raise0.3ex\hbox{$<$\kern-0.75em\raise-1.1ex\hbox{$\sim$}}}
\def\gsim{\raise0.3ex\hbox{$>$\kern-0.75em\raise-1.1ex\hbox{$\sim$}}}
\def\noi{\noindent}
\def\beq{\begin{equation}}   \def\eeq{\end{equation}}
\def\bea{\begin{eqnarray}}  \def\eea{\end{eqnarray}}

\def\noi{\noindent}
\def\di{\displaystyle}

\newcommand\mysection{\setcounter{equation}{0}\section}
\renewcommand{\theequation}{\thesection.\arabic{equation}}
\newcounter{hran} \renewcommand{\thehran}{\thesection.\arabic{hran}}

\def\bmini{\setcounter{hran}{\value{equation}}
    \refstepcounter{hran}\setcounter{equation}{0}
    \renewcommand{\theequation}{\thehran\alph{equation}}\begin{eqnarray}}

\def\bminiG#1{\setcounter{hran}{\value{equation}}
\refstepcounter{hran}\setcounter{equation}{-1}
\renewcommand{\theequation}{\thehran\alph{equation}}
\refstepcounter{equation}\label{#1}\begin{eqnarray}}

\def\emini{\end{eqnarray}\relax\setcounter{equation}{\value{hran}}\renewcommand{
\theequation}
{\thesection.\arabic{equation}}}

\begin{document}

\title {A phase-space study of the quantum Loschmidt Echo\\
 in the semiclassical limit}
\author{\it {\bf Monique Combescure} \\
\it IPNL, B\^atiment Paul Dirac \\
\it 4 rue Enrico Fermi,Universit\'e Lyon-1 \\
\it  F.69622 VILLEURBANNE Cedex, France\\
\tt email : monique.combescure@ipnl.in2p3.fr\\
\\
\it {\bf Didier Robert} \\
\it D\'epartement de Math\'ematiques\\
\it Laboratoire Jean Leray, CNRS-UMR 6629\\
\it Universit\'e de Nantes, 2 rue de la Houssini\`ere, \\
F-44322 NANTES Cedex 03, France\\
\tt email : Didier.Robert@math.univ-nantes.fr}
\vskip 1 truecm
\date{}
\maketitle

\begin{abstract}
The notion of Loschmidt echo (also called ``quantum fidelity'') has been introduced in order 
to study the (in)-stability of the quantum dynamics under perturbations of the Hamiltonian.
 It has been extensively studied in the past few years in the physics literature, in connection 
 with the problems of ``quantum chaos'', quantum computation and decoherence.\\
In this paper, we study this quantity semiclassically (as $\hbar \to 0$), taking as reference 
quantum states the usual coherent states. The latter are known to be well adapted to a 
semiclassical analysis, in particular with respect to  semiclassical estimates of their time
 evolution. For times not larger than the so-called ``Ehrenfest time''
  $C \vert \log \hbar \vert$, we are able to estimate semiclassically the Loschmidt Echo 
  as a function of $t$ (time), $\hbar$ (Planck constant), and $\delta$ (the size of the
   perturbation). The way two classical trajectories merging from the same point in classical 
   phase-space, fly apart or come close together along the evolutions governed by the
    perturbed and unperturbed Hamiltonians play a major role in this estimate.\\
We also give estimates of the ``return probability'' (again on reference states being the 
coherent states) by the same method, as a function of $t$ and $\hbar$.

\end{abstract}

\newpage

\baselineskip 18pt
\mysection{Introduction}
\hspace*{\parindent}

The semiclassical time behaviour  of quantum wavepackets has been the subject of intense interest
in the last decades, in particular in situations where there is some hyperbolicity in the corresponding
classical dynamics (Lyapunov exponents) \cite{coro1}, \cite {hajo}, \cite{ro2}. Moreover
the response of a quantum system to an external perturbation when the size $\delta$ of the
perturbation increases can manifest intriguing properties such as recurrences or decay in time of the
so-called Loschmidt Echo (or ``quantum fidelity'') \cite{co1}, \cite{co2}. 
By Loschmidt Echo we mean the following:
\\
starting from a quantum Hamiltonian $\hat H$ in $L^2(\R^d)$,  whose classical counterpart $ H$  has a chaotic 
dynamics, and adding to it a `` perturbation'' $\hat H_{\delta} = \hat H + \delta
 \hat V$, then we compare the evolutions in time  $U(t):= e^{-it\hat H/ \hbar} \quad, 
 \quad U_{\delta}(t):= e^{-it\hat H_{\delta}/ \hbar}\quad$ of initial quantum wavepackets
  $\varphi$ sufficiently well localized around some point $z$ in 
  phase-space; more precisely the {\bf overlap} between the two evolutions, or rather its
   square absolute value, is:

$$F_{\hbar, \delta}(t) := \vert \langle U_{\delta}(t)\varphi\ ,
\ U(t)\varphi\rangle \vert^2$$

For example for quantum dynamics in Hilbert space $\mathcal H = L^2(\R^d)$, $d$ being
 the space dimension, $\varphi$ can be chosen as the usual {\it coherent states},
  since they are the quantum wavepackets  ``as most  localized as possible'' in phase-space 
  $\R^{2d}$.

\bigskip

Since for $\delta =0$, we obviously have $F_{\hbar, 0}(t) \equiv 1$, and for any
 $\delta \quad, F_{\hbar,\delta}(0)= 1$, the type of decay in $t$ of
  $F_{\hbar, \delta}(t)$ so to say {\em measures the (in)fidelity of the quantum evolution 
  with respect to a perturbation of size $\delta$} for generic initial wavepackets
   $\varphi$.

The  notion of Loschmidt Echo
seems to have been first introduced by Peres (\cite{pe}),  in the following spirit: since the 
sensitivity to initial data which characterizes
     classical chaos has no quantum counterpart because of unitarity of the quantum evolution, 
     at least the ``sensitivity to perturbations'' of the Hamiltonian could replace
 it as a characterization of chaoticity in the ``quantum world''.

 \bigskip
 
A big amount of recent work appeared on the subject, studying in an essentially heuristic 
way the decay in time of $F_{\hbar, \delta}(t)$ as $t$ increases from zero to infinity; 
some of them also study this point in relationship with the important question of 
{\it decoherence}. (See \cite{becave2}, \cite{ceto2}, [12-14], [18], [22-26], [30-32]).  

In this ``jungle'' of sometimes contradictory results, it is hard to see the various arguments
 involved, in particular the precise behaviour of $F_{\hbar, \delta}(t)$ as $ \delta$ (the size
 of the perturbation), $t$ (the time),
 and of course $\hbar$ (the Planck constant)  are varied, in particular in which 
 sense and order the various limts $\delta \to 0, \hbar \to 0, t \to \infty$ are taken.\\
\noindent
Also an important point to consider is how $F_{\hbar, \delta}(t)$ depends on the location
 of the phase-space point $z$ around which the initial wavepacket $\varphi$ is peaked (since 
 classical chaoticity distinguishes various zones in phase-space with ``more or less regularity
  properties'').

\bigskip
The aim of the present paper is to start a rigorous approach of the question of semiclassical
estimate of $F_{\hbar, \delta}(t)$, in terms of classical characteristics of the (perturbed and 
unperturbed classical flows),  for initial wavepackets $\varphi = 
\varphi_{z}$  being the coherent state at phase-space point $z$. These estimates are non-perturbative,
and are carefully calculated in terms of parameters
 $(z, \delta, t, \hbar)$. The main tools we have used and developed in this
  respect are\\
1) semiclassical coherent states propagation estimates (\cite{coro1})\\
2) a beautiful formula inspired by B. Mehlig and M. Wilkinson (\cite{mewi}) about the 
Weyl symbol of a metaplectic operator, and thus of its expectation value in coherent states
 as a simple Gaussian phase-space integral ( see \cite{coro5} where we 
have completed the proof of Mehlig-Wilkinson, and treated in particular the case where the monodromy operator
 has eigenvalue 1).\\
 Note that very recently, J. Bolte and T. Schwaibold have independently obtained a  similar
  result about semiclassical estimates of the Quantum Fidelity (\cite{bolte}).

\bigskip
The plan of this paper is as follows.  In section 2 we give some preliminaries about the
Echo for suitable quantum observables, and give the semiclassics of it. In Section 3,
 we consider the (integrable) $d=1$ case,
and consider the ``return probability'' in the semiclassical limit. We give a mathematical rigorous  presentation
of   beautiful results  on ``quantum revivals'' obtained by physicists twenty years ago  (see \cite{past},  \cite{robinett}, \cite{le}). In Section 4 we consider the general 
$d$-dimensional case and give a semiclassical 
calculus of the ``return probability'' and of
   the {\bf quantum fidelity}, with precise error estimates.

\newpage

\mysection{Preliminaries}
Let us consider the quantum  Hamiltonian $\hat H_\delta = \hat H_0 + \delta\hat V$,
 depending on a real parameter $\delta$.  $\hat H_\delta$ is the Weyl quantization of  smooth classical observables defined on the phase space $\R^{2d}$. 
 Our assumptions on $H_{\delta}$ are as follows:
 \\
 {\bf Assumptions :}\\
 (H1) $H_\delta\in C^\infty(\R^{2d})$ and $\vert \partial_{X}^{\gamma}H_{\delta}(X)\vert \le C_{\gamma}, \ \forall
 X \in \mathbb R^{2d}, \ \forall \gamma : \vert \gamma \vert \ge 2$\\
 or\\
 (H'1) There exist a bounded open set $\Omega \subseteq \mathbb R^{2d}$ such that $\Omega$
 is left invariant by the classical flow $\phi_{\delta}^t,\, \forall t,\delta\in\R$
 defined by the classical Hamiltonian $H_\delta$. We assume that  the $\hbar$-Weyl
 quantization  $\hat H_{\delta}\equiv {\rm Op}_{\hbar}^{w}H_\delta$ of $H_{\delta}$
 is a self-adjoint operator in $L^2(\R^d)$  for all $\delta\in\R$. \\
 
 \noindent
 Let $L \in \mathcal C^{\infty}(\mathbb R^{2d})$ be a classical observable and
 $\hat L = {\rm Op}_{\hbar}^{w}L$. Then we assume:\\
(H2) $L \in \mathcal S(\mathbb R^{2d})$ if (H1) is satisfied,\\
(H'2)  $L \in \mathcal C_{0}^{\infty}(\Omega)$  if (H'1) is satisfied,  where $ C_{0}^{\infty}(\Omega)$
is the  linear space of $C^\infty$-smooth functions with compact support in $\Omega$.

\bigskip\noi
 Let us consider  the time evolution  unitary operator  $U_\delta(t)$, 
 in the Hilbert space $\mathcal H = L^2(\mathbb R^{d})$, 
  $$
  U_\delta(t) = \exp\left(-\frac{it}{\hbar}\hat H_\delta\right).
  $$
 
 \begin{definition}
 (i) The quantum echo is the unitary operator defined by
  \beq\label{qecho}
  E_\delta^{(q)}(t) = U_0(-t)U_\delta(t)
  \eeq
(ii) The quantum fidelity, for a state $\psi_0 $, $\Vert\psi_0\Vert = 1$, is defined by
  \beq\label{qfidel}
  f_\delta^{(q)}(t) = \left\vert\langle\psi_0 ,   E_\delta^{(q)}(t)\psi_0\rangle\right\vert^2
  \eeq
  (iii) Let $L_{\delta}(t)$ be the following time-dependent quantum observable:
  $$\hat L_{\delta}(t):= E_{\delta}^q(t)^{-1}\hat L E_{\delta}^q(t)$$
  \end{definition}
  The notion of ``fidelity'' was introduced first in classical mechanics by Loschmidt (in discussions with 
  Boltzmann) then adapted in quantum mechanics by Peres \cite{pe}. \\

   Let us define  $\phi_\delta^t$,    the classical flow defined in the phase space 
  ${\mathcal Z}\equiv\R^{2d}$ by the classical Hamiltonian $H_\delta$.  Recall  that 
  $z_{\delta,t} := \phi_\delta^t(z_0)$ is the solution  of the differential equation
  $\dot{z_t} = J\nabla H_\delta(z_t)$, $z_{t=0} = z_0$. So that the ``classical echo'' is defined by 
$$E_{\delta}^{(cl)}(t,X):= \phi_{0}^{-t}\circ \phi_{\delta}^t(X).$$
  Here $J$ is the symplectic matrix given as:
  \beq
  \label{J}
  J:= \left(
  \begin{array}{cc}
  0&\1_d\\
  -\1_d&0
  \end{array}\right)
  \eeq
  and $\1_d$ is the identity $d \times d$ matrix.\\
  \noindent
 We can see easily that in the semiclassical limit, $\hbar\rightarrow 0$, the quantum echo 
  converges to the classical echo. In more mathematical terms, the quantum echo is a $\hbar$
  - Fourier Integral Operator whose canonical relation is the classical echo.
  This is a consequence of the semiclassical Egorov theorem 
  as we shall see now, at least when the reference quantum state is a ``coherent state''. Let us recall here 
  the definition of a (Gaussian) coherent state which will be used later:\\
  Given $\varphi_{0}(x) := (\pi \hbar)^{-d/4}\exp (-x^2/ 2\hbar)$, we define, for $z:=
  (q,p) \in \mathbb R^{2d}$:
  $$
  \varphi_{z}:= \hat T(z)\varphi_{0},\;{\rm where} \; \hat T(z):= \exp\left(\frac{i(p.\hat Q-q.\hat P)}{\hbar}\right)
  $$
  are the translation Weyl operators.  
 \begin{proposition}\label{semiclass1}
 Assume either (H1-H2) or (H'1-H'2) for the Hamiltonians $H_\delta$  and observables $L$.\\
 (i) We have
 $$\lim_{\hbar \rightarrow 0}(2\pi \hbar)^d {\rm Tr}(\hat L \hat L_{\delta}(t))
 = \int_{\mathbb R^{2d}}\ L(X)L(E_{\delta}^{(cl)}(t,X))dX$$
 (ii)
  Let be $\varphi_z$ the coherent state living at $z$. Then we have:
  \beq
 \lim_{\hbar\rightarrow 0}\langle  E_\delta^{(q)}(t)\varphi_{z}\vert
   \hat L E_\delta^{(q)}(t)\varphi_{z}\rangle =  L(E_\delta^{(cl)}(t,z))
   \eeq
   Moreover there exists $C>0$ such that the limits are uniform as long as
    $\vert t \vert \le C \vert \log \hbar \vert$. If $H_{0}$ is integrable in $\Omega$
    and if $\delta = {\mathcal O}(\hbar)$, then the limit is uniform as long as  $\vert t \vert \le C_{\varepsilon}
    \hbar^{-1/3 + \varepsilon}$
   \end{proposition}
   
   {\bf Proof}: It follows from the semiclassical Egorov Theorem,with improvement for large times
   derived by Bouzouina-Robert (\cite{boro}).
   \medskip

     \medskip
     
       \begin{remark} An important question is to control the time of validity of the 
      semiclassical approximation. Rigorous mathematical results are far from  numerical and 
      theoretical expected physical results. Without assumptions on classical flows this time is 
      the Ehrenfest time (of order
       $\log(\hbar^{-1}))$. 
        \end{remark}
          
   \mysection{Revivals for 1-D systems}   
   
   In this Section we shall consider the Return Probability, which is a simplified form of Quantum
   Fidelity as we shall explain in the next Section. For one-dimensional problems, this
   Return Probability  manifests interesting recurrences very close to 1, as time evolves.
   This phenomenon  was studied in the physics literature to understand time evolution of Rydberg
   atoms  and their quantum beats,  with decay and reformation of the wave packet (see for example \cite{past,le},
   \cite{robinett} and references herein contained).
      In this section we want to give a flavour of  results  obtained by physicists in the last twenty years, concerning  revivals for the  quantum return probability (\cite{robinett} for a very clear and detailed review) 
 and show how to put them in a more  rigorous mathematical framework.
   Let us consider a classical 1-D Hamiltonian $H$. One assumes $H$ to be a  smooth, confining 
    with one well Hamiltonian. This  means that the energy surface   $H^{-1}(E)$  
          has   only one connected component in phase-space ${\mathcal Z}.$
   Let  $\Psi_n$ be an orthonormal basis of eigenstates, with eigenvalues $E_n$, 
   $n\in\N$.\\
   Let  $\di{\psi_0 = \sum_{n\in\N}c_n\Psi_n}$ an initial  normalized state,
    and $\psi_t = U(t)\psi_0$. Then the  autocorrelation fonction is :
   \beq\label{autocorr}
   a(t): = 
  \langle\psi_0\vert\psi_t \rangle = \sum_{n\in\N}\vert c_n\vert^2{\rm e}^{-\frac{it}
  {\hbar}E_n}
  \eeq
   and the return probability is defined by 
   \beq
   \rho(t) = \vert a(t)\vert^2.
   \eeq
    Let us remark  here that $a$ is  an almost periodic function  (in the sense of H. Bohr)  in time $t$ on $\R$.
     Therefore, for every $\varepsilon >0$,  there exists $T_\varepsilon>0$  and for every $k\in\Z$ there exists 
     $t_k\in[kT_\varepsilon, (k+1)T_\varepsilon[$ such that $\vert a(t_k)-1\vert \leq\varepsilon$.
    This fact can be  interpreted as  a   quantum analog  of the famous return  Poincar\'e in classical mechanics.\\
   But we have no information  here on the almost return time $t_k$, in particular when $\hbar$ tends to zero.
     For 1-D systems much more accurate results are available because for these systems
      the spectrum can be computed with error ${\mathcal O}(\hbar^\infty)$  according the Bohr-Sommerfeld quantization rule. 
      Recall here this result.
      We take the presentation from the paper by Helffer-Robert (\cite{hero})
       and we refer to this paper for more details.  (see also the thesis of Bily for a proof  using  coherent states). 
       \\ Let us give now the  sufficient assumptions:\\
        $(A_1)$  $H(z)$ {\em is real valued },   $H\in C^{\infty}({\mathcal \R^2})$. \\
$(A_2)$ $H$ is bounded below\footnote { Using the semi-classical functional calculus \cite{ro1}
it is not a serious restriction} ~: there exist  $c_0>0$
 and $\gamma_0\in\R$ such that $c_0\leq H(z) +\gamma_0$. Furthermore 
 $H(z) +\gamma_0$ is supposed to be  a temperate weight, i.e
{\em there exist $C>0,\; M\in\R$,  such that}~:
$$
H(z)+\lambda_0 \leq C(H(z^\prime)+\lambda_0)(1+\vert z-z^\prime\vert )^M\;\;\;
\forall z, z^\prime  \in Z.
$$
$(A_3)$ $\forall \gamma$ {\em multiindex  
 $\exists c>0 $ such that}: 
$\vert \partial^\gamma_z H\vert \leq c(H+\lambda_0)$.\\

We want to consider here bound states of $\hat H$  in a fixed energy band.  So, let us consider a  classical energy 
 interval $I = ]E_--\varepsilon , E_++\varepsilon[, \;\; E_- < E_+$ such that we have:\\
$(A_4)$ 
 $H^{-1}(I) $ {\em is a bounded set of the phase space} 
$\R^{2}$.\\
This implies that in the closed  interval $ I = [E_-, E_+]$, 
 for $\hbar > 0$ small enough, the spectrum of $\hat{H}$ in $I$
is purely discrete (\cite{hero}).\\ \noi
For some energy level $E\in I$, let us introduce the assumption                   :\\ 
 $(A_5)$  $E$ {\em is a regular value of $H$. That means}:
 $H(x,\xi)= E \Rightarrow \nabla_{(x,\xi)}H(x,\xi) \neq 0$.\\
 Furthermore we assume that   for every $E\in I$, $H^{-1}(E)$ is a connected curve.\\
      Let us consider a non critical  energy interval $[E^-, E^+]$. It is well konwn that the action integral is
       ${\mathcal J}(E) = \int_{H(z)\leq E}dz$ and the period along the energy curve $H^{-1}(E)$
        is $T_E = {\mathcal J}^\prime(E)$, $E\in [E^-, E^+].$          
         Let us denote $F^\pm = {\mathcal J}(E^\pm)$.  The eigenvalues of $\hat H$
          in $[E^-, E^+]$ are determined  by the following Bohr-Sommerfeld rule.

          \begin{theorem}[\cite{hero}] Under the assumptions $(A_1)$ to $(A_5)$,  there exists
           a $C^\infty$ function on $[F^-, F^+]$,  $F\mapsto 
          b(F,\hbar)$ and
           $C^\infty$ functions $b_j$  defined on $[F^-, F^+]$ such that 
           $\di{b(F, \hbar) = \sum_{j\in\N}b_j(F)\hbar^j + {\mathcal  O}(\hbar^\infty)}$ and
            the eigenvalues  $E_n$ of $\hat H$ in $I$  are   given by 
            \bea\label{bs}
            E_n = b((n+\frac{1}{2})\hbar, \hbar) +{\mathcal O}(\hbar^\infty),\;
            {\rm for}\; n\;{\rm such\; that}\;( n+\frac{1}{2})\hbar\in [F^-,F^+] \\
            {\rm where}\; b_0(F) = 2\pi{\mathcal J}^{-1}(F),\;{\rm and}\;\boxed{ b_1= 0}.
            \eea
            \end{theorem}
            \begin{remark}
             In  recent papers   \cite{cdv}, \cite{lit} the authors  have  given  some 
           methods to compute  explicitly the
              terms $b_j$ for $j\geq 2$ in the expansion in $\hbar$  in  the Bohr-Sommerfeld rule. 
             \end{remark}
                         Let us now choose an initial  wave packet,
                          $\di{\psi = \sum_{n}c_n\Psi_n}$,  tightly spread around the energy $E_{\bar n}$
                     where  $E_{\bar n} = b(\bar n +\frac{1}{2}, \hbar)$ and
            $\bar n$  is a family of  given quantum numbers, depending on $\hbar$,
             and such that $(\bar n+\frac{1}{2})\hbar\in [F^-, F^+]$  for every $\hbar\in]0, \hbar_0]$.  
             Let us choose the coefficient $c_n$ defined by 
               $$
               c_n = K_{\tau,\hbar}\chi_1\left(\frac{E_n-E_{\bar n}}{\tau_\hbar}\right)
               \chi_0\left(\frac{E_n-E}{\epsilon_\hbar}\right)
               $$
               where $\chi_ 1\in{\mathcal S}(\R)$, $\chi_0$ has a bounded support,
                supp$[\chi_0]\subseteq ]-1, 1[$, $\chi_0(x)=1$ on $[-1/2, 1/2]$,   
                and $ K_{\tau,\hbar}$ is defined such   that the $L^2$-norm of the  wave packet is 
                $\di{\sum_{n\in\N}\vert c_n\vert^2 = 1}$.  We shall choose $\tau_\hbar$ and $\epsilon_\hbar$
                 such that $\frac{\tau_\hbar}{\epsilon_\hbar} = {\mathcal O}(\hbar^{\delta})$, for some $\delta>0.$
                \begin{remark}
               From a physical point of view,  a state $\psi$ as above is prepared by exciting an atom with a laser beam.
               The  new object is a Rydberg atom.
                \end{remark}

                Pratically, we shall choose $\tau_\hbar = \hbar^\theta$. We define: $\sigma \equiv \frac{\tau}{\hbar} =\hbar^{\theta-1}$
                and 
                $\epsilon_\hbar = \hbar^{\theta^\prime}$, 
                 with  $0< \theta^\prime<\theta<1$. It is more suggestive for us  to keep the notations $\tau_\hbar$
                  $\sigma_bar$ and $\epsilon_\hbar.$\\
                   Let us apply the Taylor formula to $b(F,\hbar)$ around $ \bar F = (\bar n+\frac{1}{2})\hbar$. 
                                \beq\label{tayl}
             E_n - E_{\bar n} = \hbar b_0^\prime(n-\bar n) + 
             \frac{\hbar^2}{2}b_0^{\prime\prime}(n-\bar n)^2 + \frac{\hbar^3}{6}b_0^{\prime\prime\prime}
             (n-\bar n)^3 +  \hbar^3b^\prime_2(n-\bar n) + {\mathcal O}(\tau^4),
             \eeq
             if $\vert n-\bar n\vert \leq C\sigma$, with $C>0$ and 
              where the derivatives of $b_j$ in $ F$ are computed in
               $\bar F.$
               Up to a small error in $\hbar$,  it is  possible  to change the definitions of
                 $\chi_1$ and $ K_{\tau,\hbar}$
                 such that $c_n =  K_{\tau,\hbar}\chi\left(\frac{n-\bar n}{\sigma}\right)$, with
                  $\sigma = \frac{\tau}{\hbar}$. Let us remark that 
                $ K_{\tau,\hbar}$ is of order $ \sigma_\hbar^{-1}$. This is  easily seen from the following lemma.
                \begin{lemma}
                With the  previous notations and assumptions we have
                \beq
                \lim_{\hbar\rightarrow 0}\left(\frac{\hbar}{\tau}\sum_{n}
                \left\lvert\chi_1\left(\frac{E_n-E_{\bar n}}{\tau_\hbar}\right) \chi_0\left(\frac{E_n-E}{\epsilon_\hbar}\right)\right\rvert^2\right)     = \int_{\R}\vert\chi_1(x)\vert^2dx
              \eeq
                \end{lemma}
                {\em Proof}. Besides the assumptions,  we make use of  formula (\ref{tayl})
                and of the following well known estimate for the number of bound states
                $$
                \#\{n,\; E_n\in[E-\epsilon, E+\epsilon]\} = {\mathcal O}\left(\frac{\epsilon}{\hbar}\right).
                $$
                The details are left to the reader.
               \sq

                                   Let us denote by $a_i(t)$    the approximation for
                     $a(t)$ obtained by plugging in (\ref{autocorr}) the i-first terms 
                       of the Taylor expansion (\ref{tayl}) denoted by
                       $\kappa_i(n)$ ($1\leq i\leq 3$).  So we get the  following preliminary result:
                             \begin{proposition}
                       We have
                       \beq\label{autoapprox}
                       \vert a(t)\vert^2 =
                        \vert a_i(t)\vert^2 + {\mathcal O}(\vert t\vert\hbar^{-1}\tau^{i+1})
                        \eeq
                         In particular, 
                            $\vert a_i(t)\vert^2$ is a  semiclassical approximation  for 
                           $ \vert a(t)\vert^2$  valid for times t such that  $\vert t\vert$ is
                        less than $\hbar^{1+\varepsilon}\tau^{-1-i}$   for any $\varepsilon>0$,
                         with a reminder term ${\mathcal O}(\hbar^\varepsilon)$.
       \end{proposition}
       \begin{corollary}
       For every $\varepsilon >0$, we can choose $\theta<1$, close enough to 1, such that
     \beq
     \vert a(t)\vert^2 =
                        \vert a_i(t)\vert^2 + {\mathcal O}(\vert t\vert\hbar^{i-\varepsilon})
             \eeq
       \end{corollary}
                      From the proposition and its   corollary   we can give a mathematical  proof 
                      for  the collapses and  revivals phenomenon
                      concerning the return probability $\rho(t)$. \\
                             Let us remark  first that  $\kappa_1(n) = b_0^\prime (n-\bar n)$ so 
                      $\vert a_1(t)\vert^2$ is periodic with period $T_{cl}= \frac{2\pi}
                      {b_0^\prime}$  
                       (classical period along the orbit of energy $\bar E$). 
                        So the return probability $\rho(t)$  is close to 1 
                        for $t = NT_{cl}$ as far as $\vert t\vert$ is less than
                         $\tau^{-2}\hbar^{1+ \varepsilon} = {\mathcal O}(\hbar^{-1+\varepsilon^\prime})$
                         ($\varepsilon^\prime>\varepsilon.$)\\ For larger times, 
                       we have  to consider  $\kappa_2(n) = b_0^\prime (n-\bar n) + \frac{\hbar}{2}
                       b_0^{\prime\prime}(n-\bar n)^2$ and 
                      a second time scale dependent on $\hbar$, the revival time,  defined as 
                      $T_{rev} = \frac{4\pi}{\hbar b_0^{\prime\prime}}$.  
                      Let us introduce the integer $N =\left\lbrack\frac{T_{Rev}}{T_c\ell}\right\rbrack$ and take 
                      $NT_{c\ell}$ as a new time origin. If $t = NT_{c\ell} + s$ we have
                      $$
                      a_2(t) = \sum\vert c_{m+\bar n}\vert^2\exp\left(\frac{2i\pi}{T_{c\ell}}sm\right)
                      \exp\left(\frac{2i\pi}{T_{Rev}}(s-\theta T_{c\ell})m\right),
                      $$
                      where $\theta\in[0, 1[$. Therefore we have
                      \beq 
                      a(NT_{c\ell}+s) = a(s) + {\mathcal O}(\hbar^\varepsilon),\;
                      {\rm as\;long\; as}\;  \vert s\vert \leq \hbar^{\varepsilon-2}
                      \eeq
                      So, around the time  $NT_{c\ell}$, which is of order $\hbar^{-1}$, the signal  retains its initial form and moves according to 
                      the classical laws.\\
                                              Now we  shall prove  that for large time intervals, below the time   
                                               $T_{Rev}$ the signal  $a(t)$ is very small. Let us consider
                       $$
                       J_{\hbar} = 
                       [\hbar^{1-2\theta-\delta_1}, \hbar^{\delta_2/2-\theta}],
                       $$
                       where $\delta_1 >0$, $\delta_2>0$ are any small fixed real numbers satisfying
                        $\delta_2+\delta_1/2+\theta < 1$.
                        \begin{proposition}
                        Under the previous assumptions and notations,  we have
                                      \beq\label{collapse}
                       \lim_{\hbar\rightarrow 0, t\in J_{\hbar}}\rho(t) = 0
                       \eeq
                         \end{proposition}
                         {\em Proof}: 
                       For  simplicity, we shall prove  the collapse property (\ref{collapse}) for a Gaussian  cut-off, 
                        $\chi_1(x) = {\rm e}^{-x^2/4}$. We can assume that $T_{c\ell} = 2\pi$. 
                         \\
                         The trick here is to apply the Poisson formula in the time variable  to
                         \beq
                         a_2(t) = 
                         K_{\tau,\hbar}\sum_{m\in\Z}\exp\left(-  \frac{m^2}{2\sigma^2} + 2i\pi t\frac{m^2}{T_{rev}}\right)
                          \exp(itm)
                          \eeq
                         So, applying the classical formula for the Fourier transform of a  Gaussian we get
                         \beq\label{Poisson}
                         a_2(t) = K_{\tau,\hbar}\sqrt{\frac{2\pi}{\gamma_{t,\hbar}}}\sum_{\ell\in\Z}
                         \exp\left(-2\pi^2\frac{(\ell - \frac{2\pi t}{2\pi})^2}{\gamma_{t,\hbar}}\right)
                         \eeq
                         where $\gamma_{t,\hbar} = \left(\frac{1}{\sigma^2} - \frac{4i\pi t}{T_{rev}}\right)$.\\
                         We have
                         $$
                         \gamma_{t,\hbar}  = \gamma_{0,\hbar} \left(1  -  \frac{4i\pi t}{T_{rev}}\sigma^2\right)
                         $$
                         and each Gaussian term  in the sum in (\ref{Poisson}) has width $\delta_t$, given by
                         $$
                         \delta_t = \left(\Re({\gamma_{t,\hbar}}^{-1})\right))^{-1/2} = 
                         \left(\frac{1}{\sigma^2} + 16\pi^2\frac{t^2\sigma^2}{{T_{rev}}^2}\right)^{1/2}
                         $$
                        From formula (\ref{Poisson}), we can see that a sufficient condition for $t$
                         to be a collapse time for $\rho(t)$ is that 
                        $\vert\frac{\gamma_{0,\hbar}}{\gamma_{t,\hbar}}\vert$ and  $\delta_t$ tend to 0 with $\hbar$. 
                        Therefore we get easily (\ref{collapse}).\sq
                        \begin{remark}
                        The length of $J_\hbar$ is of order 
             is of order $\hbar^{\delta_2/2-\theta}$. So the  length of $J_\hbar$
                         is very large for $\hbar$ very small (remember that $\delta_2$ is small and $\theta$ close to 1). 
                         Therefore   in the large  intervall $J_\hbar,$ $a(t)$ is very small 
                         and  in particular its classical period $T_{c\ell}$  has disappeared.  But we have seen that this period  appears again
                         at time $NT_{c\ell}$ (close to $T_{Rev}$,  for small $\hbar$.
                         These facts  justify the name ``revival"  given to this phenomenon.\\
                          As it is shown in \cite{robinett}, it is also possible to observe fractional revivals, using some elementary  properties of integers.
                        \end{remark}
                        
                              \begin{remark}
                       The above analysis could be extended to completely integrable systems
                        in $d$ degree of freedom, using the corresponding Bohr-Sommerfeld rules \cite{char}.
             \end{remark}
                       
                     \begin{remark}
                     In the next section, for $d$-multidimensional  sytems,  we shall start with a Gaussian coherent 
                     $\varphi_z$ of classical energy $E = H(z)$. Let us consider $\chi$ as above and such that $\chi =1$
                      in a small neighborhood of $E$. Then, modulo an error term ${\mathcal O}(\hbar^\infty)$,
                      we have easily
                      \beq\label{1Dcoh}
                      \langle \varphi_z\vert U(t)\varphi_z\rangle = \sum_{n\in\N}
                      \chi\left(\frac{E_n- E}{\tau}\right)\vert\langle\varphi_z\vert e_n\rangle\vert^2
                      {\rm e}^{-\frac{it} {\hbar}E_n}
                      \eeq
                      We get something similar to the definition of $a(t)$ but with  coefficients $c_n$ not necessary
                       smooth in the variable $n$, so application of the Poisson formula seems difficult.
                     \end{remark}

   \mysection{Fidelity on coherent states}
   
   Let us recall that the Return Probability for suitable time-dependent Hamiltonians
   $\hat H(t)$ (for which the quantum unitary evolution $U(t,s)$ can be shown to exist)
   in  some quantum state $\psi_{0} \in \mathcal H = L^2(\mathbb R^d)$ is defined as
   \beq
   \label{return}
   R(t):= \vert \langle\psi_{0}, U(t,0)\psi_{0} \rangle \vert^2
   \eeq
   It measures the quantum probability that the time-evolved quantum state $U(t,0)\psi_{0}$
   returns close to its initial quantum configuration $\psi_{0}$.\\
   The Quantum Fidelity (\ref{qfidel}) can be related to the Return Probability for a suitable
   time-dependent Hamiltonian:
   \beq
   \label{timedep}
   \hat H(t)= \exp\left(\frac{it\hat H_{0}}{\hbar}\right)(\hat H_{\delta}
   -\hat H_{0})\exp\left(-\frac{it\hat H_{0}}{\hbar}\right)
   \eeq
   Namely, according to Schr\" odinger equation, we have that, defining 
   $U(t):= U_{0}(-t)U_{\delta}(t)$, 
   $$i\hbar \frac{d}{dt}U(t)=\hat H(t)U(t)$$
   so that $U(t,s)= U(t)U(-s)$ is the time evolution associated to (\ref{timedep}).\\
   
   \noindent
   Thus as a training for studying Quantum Fidelities, let us first consider the semiclassical
   study of the Return Probability in the coherent states.\\
   
   \noindent
   Let us assume the following hypotheses:\\
    \begin{enumerate}
                       \item $H(t,X)$  is a smooth time dependent Hamiltonian, continuous in time $t\in\R$,
                        $C^\infty$ in $X\in\R^{2d}$  such that   for every multiindex 
$\alpha$ there exist $C_\alpha > 0$ and $M_\alpha \in\R$
 such that
\beq\label{CS}
\vert\partial_X^\alpha H(t,X)\vert \leq C_\alpha (1 + \vert X\vert)^{M_\alpha},
\;\; {\rm for}\;  X\in\R^{2d}, t\in\R. 
\eeq
\item The classical flow $\phi^{t,s}$ generated by $H(t)$ (with initial data at $s$) exists for all times $t,s$.
We shall denote $\phi^t=\phi^{t,0}$.
     \item $\hat H(t):= {\rm Op}_{\hbar}^w H(t)$ is self-adjoint on $L^2(\R^d)$, and 
     generates a strongly continuous evolution operator $U(t,s)$ satisfying the chain rule
     $U(t,\tau)U(\tau,s)= U(t,s), \ \forall s, \tau, t \in \mathbb R$.\\
     Note that sufficient conditions for this to hold are given in \cite{coro1}.
     \end{enumerate}
       \noindent
     Then we define the Stability Matrix $F(t)$ for the flow $\phi^{t}$ as follows:\\
     It is  the $2d \times 2d$ symplectic matrix solution of the following linear problem:
     $$\dot F_{t}= J H''(t)F_{t}$$
     where $H''(t)$ is the Hessian of $H(t)$ taken at point $\phi^{t}z$ of the phase-space
     trajectory, starting with initial phase-space point $z \in \mathbb R^{2d}$, $J$ being
     the symplectic matrix given by (\ref{J}). We have:
     
     \begin{theorem}
  Let us assume Hypotheses 1,2,3 above. Then  we have, for the amplitude of the return probability
 $r(t,z):= \vert \langle U(t,0)\varphi_{z} ,  \varphi_{z}\rangle\vert$, the asymptotic 
     formula
      as $\hbar\rightarrow 0$,
           \beq
     r(t,z) = 
     \vert{\rm det}(V_t)\vert^{-1/2}{\rm e}^{\frac{\Re\triangle_t}{\hbar}} + {\mathcal O}(\sqrt\hbar)
     \eeq
     where 
     $$V_{t}=\frac{1}{2}(\1+F_t+iJ(\1-F_t)) $$
     $F_t$ being the stability matrix for the flow, and
      $$
     \triangle_t = \frac{1}{4}\Gamma_{F_t}(z_{t}- z)\cdot (z_{t}-z)
     $$ 
     with
 $\Gamma_{F_t} = (\1+iJ)(\1+F_t)(2V_{t})^{-1}(\1-iJ) - \1$
 \\
 In particular if $z$ lies on a periodic orbit $\gamma$ of the classical flow, with period 
 $ T_{\gamma}$,
 and if $F(T_{\gamma})$ is unitary, we get:
 $$r(T_{\gamma}, z)= 1+ {\mathcal O}(\hbar^{1/2})$$
 namely we have  almost ``quantum revival'' when $\hbar \to 0$.     
 \end{theorem}      
 
 The proof will be very similar to the one we establish below for the Quantum Fidelity.\\
 
 \noindent
 Let us now consider the fidelity problem. 
 We want to analyze
 $$f_{z}^{\delta}(t):= \vert \langle U_{0}(t)\varphi_{z}\ ,\ U_{\delta}(t)\varphi_{z}
 \rangle \vert^2.$$
 For the generators of the (time independent) Hamiltonians $\hat H_{\delta}$, 
 we assume:\\
 
 (H0) $\hat H_{\delta}= {\rm Op}_{\hbar}^w H_{\delta}$\\
 
 (H2) $H_{\delta}$ is a smooth Hamiltonian such that there exists for any multiindex
 $\gamma$ constants $C_{\gamma}>0$, and $m_{\gamma }\in \mathbb R$ such that
 $\vert \partial_{X}^{\gamma}H_{\delta}(X)\vert \le C_{\gamma}(1+ \vert 
 X \vert)^{m_{\gamma}},\ \forall X \in \mathbb R^{2d}$.
 
 We denote by $F_{\delta}(t)$ the stability matrix for $H_{\delta}$ (and similarly for
 $F_{0}(t)$ and $H_{0}$), and by $\phi_{\delta}(t)$ the classical flow for $H_{\delta}$,
 so that the phase-space point of the classical trajectory starting from $z \in \mathbb R^{2d}$
 is $z_{t}^{\delta}\equiv \phi_{\delta}^tz$.\\
 
 \noindent
 Then we have:
 
 \begin{theorem}
Assume (H0), (H2). \\
(i) Then for any $N \ge 1$ we have the asymptotic expansion:
\beq
\label{exponent}
\langle U_{0}(t)\varphi_{z}\ ,\ U_{\delta}(t)\varphi_{z}\rangle=
\sum_{j=0}^N \hbar^{j/2}e_{j}\left(t,z,\frac{z_{t}^{\delta}-z_{t}^0}{\sqrt \hbar}
\right)+R_{\delta}^{(N)}(t, \hbar)
\eeq
where
$$\bullet\ e_{j}(t,z,X):= \alpha_{j}(t,z,X)e^{\Lambda_{t,z}X\cdot X}$$
$$\bullet\ \alpha_{j}(t,z, .)\ \mbox{is a polynomial of degree}\ \le 3j$$
$$\bullet\ \Lambda_{t,z}:= \frac{1}{4}\tilde F_{0}^{-1}\Gamma_{F}F_{0}^{-1},
\ \tilde F \ \mbox{being the transpose of}\ F$$
$$\bullet\ \Gamma_{F}:= (\1+iJ)(\1+F)\left(\1+F+iJ(\1-F)\right)^{-1}(\1-iJ)-\1$$
$$\bullet\ \mbox{and}\ F \ \mbox{denotes}\ F:= F_{0}^{-1}F_{\delta}$$
$R_\delta^{(N)}(t,\hbar) = {\mathcal O}(\hbar^{(N+1)/2}) $ is uniform on every interval
$ [-T, T]\  (0<T<\infty)$.\\
 In particular we have:
$$e_{0}(t,z)\equiv \left(\det\left(\frac{1}{2}(\1+F+iJ(\1-F))\right)\right)^
{-1/2}$$
and denoting by $V_{F}$ the following $2d\times 2d$ matrix:
\beq
\label{VF}
V_{F}:= \frac{1}{2}(\1+F+iJ(\1-F))
\eeq
$$f_{z}^{\delta}(t)= \vert \det V_{F}\vert^{-1}\exp\left(\frac{2}{\hbar}
\Re\Lambda_{t,z}(z_{t}^{\delta}-z_{t}^0)\cdot (z_{t}^{\delta}-z_{t}^0)\right)
+{\mathcal O}(\sqrt \hbar)$$
(ii) Moreover, we have, in the sense of quadratic forms the following inequality:
\beq
\label{Delta}
\Re \Lambda_{t,z}\le -\frac{1}{2+2\Vert F\Vert^2}\tilde F_{0}^{-1}F_{0}^{-1}
\eeq
where $\Vert F \Vert$ is the largest eigenvalue of $F$, and for any symplectic matrix $F$:
$$\vert \det V_{F}\vert \ge 1$$
(iii) $\di{\lim_{\hbar \to 0}f_{z}^{\delta}(t)=1}\ \iff\ z_{t}^{\delta}= z_{t}^0\ 
\mbox{and}\ F\equiv F_{0}^{-1}F_{\delta}$ is a unitary matrix.
 \end{theorem}                

\begin{theorem}
Let us assume that $H_{\delta}(t)$ obeys a Gevrey class assumption:\\
$(G_{s}):\  \exists \rho>0, \nu \ge 0, C>0, R>0$ such that:
$$\forall \gamma \in \mathbb N^{2d},\ \left\vert \partial_{X}^{\gamma}H_{\delta}(X)
\right\vert \le C R^{1+\vert \gamma\vert}(\gamma!)^s e^{\nu\vert X \vert^{1/s}},\ 
\forall X \in \mathbb C^{2d},\ \vert \Im X \vert \le \rho,\ \forall \delta 
\in[0,\delta_{0}],\ \delta_{0}>0$$
then with the notations of Theorem 4.2, we have:\\
$(i)$  $\vert e_{j}(t,z,X)\vert\le C^{j+1}(j+1)^{\frac{s_\star(j+1)}{2}}e^{-\mu\vert X\vert}$,
 where $s_\star = 2s-1$.
 \\
$(ii)$  $\forall\varepsilon>0$  $\exists a>0$, $\exists c>0$ such that  for $\vert X\vert\geq\varepsilon$
  and $j\leq \frac{a}{\hbar^{1/s_\star}}$ 
 we have $\vert e_{j}(t,z,X)\vert \leq {\rm e}^{-\frac{c}{\hbar^{1/s_\star}}}$.\\
 $(iii)$ $\exists C>0,\ c>0$ such that if $N_{\hbar}= \left[\frac{a}{\hbar^{1/s_\star}}\right]$
we have $\forall t : \ \vert t\vert\le T$:
$$\left\vert \langle U_{0}(t)\varphi_{z}\ ,\ U_{\delta}(t)\varphi_{z}\rangle
-\sum_{j=0}^{N_{\hbar}}\hbar^{j/2}e_{j}\left(t,z, \frac{z_{t}^{\delta}-z_{t}^0}{\sqrt \hbar}
\right)\right\vert \le C\exp\left(-\frac{c}{\hbar^{1/s_\star}}\right)$$
\end{theorem}

The proof of these theorems heavily relies on a result for semiclassical propagation of
coherent states (see \cite{coro1}) which has been revisited in \cite{ro2}, \cite{ro3}:

 \begin{theorem}\label{gene} Assume (H0), (H2). Then  
  there exists a family of polynomials   $\{b_j(t, x)\}_{j\in \mathbb N}$ 
in $d$ real variables $ x= (x_1,\cdots, x_d)$,
    with time dependent coefficients, such that for all $\hbar\in ]0, 1]$, 
  we have
\beq\label{propag1}
\left\Vert U(t)\varphi_{z} - 
 \exp\left(\frac{i \gamma_t}{\hbar}\right){\hat T} (z_t) \Lambda_\hbar 
 \widehat{R_1}(F_t)
 \left(\sum_{0\leq j\leq N}\hbar^{j/2}b_j(t)g\right) \right\Vert_{L^2(\R^d)}
  \leq 
 C(N, t, \hbar)\hbar^{(N +1)/2}
 \eeq
 such that for every $N\in \mathbb N$, and every  $T< +\infty$ we have  
 $\displaystyle{\sup_{0< \hbar \leq 1,\vert t\vert\leq T} 
 C(N, t, z, \hbar) < +\infty}$. 
  $g$ is the simple normalized gaussian function $\mathbb R^d\mapsto \mathbb R$:
     $$g(x):= (\pi)^{-d/4}\exp\left(-\frac{x^2}{2}\right)$$ 
 Here  $\widehat R_{1}(F)$ is the usual metaplectic 
 representation     (for $\hbar=1$) associated to $F$ (see \cite{coro5}). Moreover
 $\Lambda_{\hbar}$ is the following unitary transform in $\mathcal H$:
 $$
      \Lambda_{\hbar}\psi(x) = \hbar^{-d/4}\psi\left(x\hbar^{-1/2}\right)  \;\;
{\rm  and}\;\;$$
$$\gamma_t(z) = \frac{1}{2}\int_0^tz_s\cdot\nabla H(z_s)ds - tH(z)$$
\end{theorem}

Let us here recall a simple property of the metaplectic representation:\\
if  $$F = \left(
     \begin{array}{cc}
     A&B\\
     C&D
     \end{array}\right)$$
     is the 4 $\ d \times d$ block-matrix form of the symplectic matrix $F$, the action of $\widehat R_{1}(F)$
     on the state $g$ is given by:
     $$\widehat R_{1}(F)g = \pi^{-d/4}(\det (A+iB))^{-1/2}\exp \left(\frac{i}{2}
     \Gamma x \cdot x\right)$$
     with $\Gamma := (C+iD)(A+iB)^{-1}$
     
     \bigskip
        \noindent
         Let us denote by $\psi_{z,t}^{(N)}$ the approximation of $U(t)\varphi_z$
         given  by (\ref{propag1}). \\
     Let us recall some more accurate estimate obtained in \cite{coro1} and \cite{ro2, ro3}.\\
      \begin{enumerate}
    \item[(i)] Let be $N$ fixed and $R > 0$ such that $\vert z_t\vert\leq R$, $\forall t
    \in\R$.
    Then there exist $c_N>0$, $k_R>0$ such that
        \beq\label{rest1}
          \hbar^{(N+1)/2}C(N, t, z, \hbar) \leq c_Nk_R\left(\sqrt \hbar\vert F_{t}
          \vert^3\right)^{N+1}(1+\vert t\vert)^{N+1}
                                     \eeq
    In particular, in the generic case,  we have a positive Lyapunov exponent $\gamma$ such that
     $\vert F_{t}\vert \leq {\rm e}^{\gamma\vert t\vert}$, so that the semiclassical 
   approximation is valid for $\vert t\vert \leq \frac{1-\varepsilon}{6\gamma}\vert 
   \log\hbar\vert$.\\
   
  In the integrable case we have $\vert F_{t}\vert \leq c\vert t\vert$ and 
                                       the semiclassical approximation is valid for 
         $\vert t\vert \leq \hbar^{-1/6+\varepsilon}$, for any $\varepsilon >0$.
         
   \item[(ii)] If $H$ satisfies the following analyticity assumption in the set 
                                       \beq
\Omega_\rho = \{X\in \C^{2n}, \vert\Im X\vert < \rho\}
\eeq
where $\Im X = (\Im X_1,\cdots, \Im X_{2d})$ and $\vert \cdot \vert$ is
 the Euclidean norm in $\R^{2d}$ for the Hermitean norm in $\C^{2d}$. So we assume
  there exist 
 $\rho>0$,  $C>0$, $\nu \geq 0$,  such that  $H$ is holomorphic in $\Omega_\rho$
 and  for all  $X\in \Omega_\rho$,   we have 
\beq\label{analyt}
\vert H(X)\vert \leq C{\rm e}^{\nu\vert X\vert}.
\eeq
      Then the $N$-dependent constant $c_N$ in (\ref{rest1}) can be estimated by
      \beq
      c_N \leq C^{N+1}(N+1)^{\frac{N+1}{2}}
      \eeq 
      From this estimate we get an approximation for $U(t)\varphi_z$ modulo an 
      exponentially small error
       (see also \cite{hajo}).
       
       \item[(iii)] There exist $\tau>0$, $a>0$, $k>0$ such that for 
       $N = \{\frac{a}{\hbar}\}$ (the nearest integer to
        $\frac{a}{\hbar}$), we have
        \beq
        \left\lVert U(t)\varphi_z - \psi_{z,t}^{(N)}\right\rVert \leq k{\rm e}^{-\frac{\tau}{\hbar}},
        \; \forall \hbar\in ]0, 1].
        \eeq
           \end{enumerate}
             Now we apply the above estimates and the results already proven \cite{coro5} 
             concerning  the  action of metaplectic transformations on Gaussians. 
              Our aim is to study the fidelity
               
              \beq
              \label{qf}
              f_{\delta, z}(t) =  \vert\langle U_0(t)\varphi_z , U_\delta(t)
              \varphi_z\rangle\vert^2
              \eeq
               
     We shall add  the index  $\delta$ to keep  track of  the dependence on the perturbation
               parameter in the Hamiltonian $H_\delta$.
                $z$ is fixed so we shall omit index $z$.\\ 
                We use the approximants  $\psi_{z,t,\delta}^{(N)}$ 
                and $\psi_{z,t,0}^{(N)}$ for both terms of the  scalar
                product in (\ref{qf}). This yields that to get the result of Theorem 4.2 mod 
                ${\mathcal O}(\sqrt \hbar)$ we have to calculate:
     \beq
     \label{qf1}
     \left\langle \hat T^{(\hbar=1)}\left(\frac{z_{t}^0-z_{t}^\delta}{\sqrt \hbar}
     \right)\hat R^1(F_{t}^0)g\ ,\ \hat R^1(F_{t}^{\delta})g\right\rangle
     \eeq
    
     But (\ref{qf1}) is simply of the form
     $$e^{i\beta_{t}/\hbar}\langle\hat T^{(\hbar=1)}\left(F_{t}^0)^{-1}\left(\frac{z_t^0-z_t^\delta}
     {\sqrt \hbar}\right)\right)g\ ,\ \hat R^1((F_{t}^0)^{-1}F_{t}^{\delta})g \rangle$$
     where
     $$\beta_{t}:= -\frac{1}{2}\sigma(z_{t}^{\delta}
     ,z_{t}^0).$$   
       Recall that $\sigma(X,Y)\equiv X\cdot JY$ is the symplectic form in
            $\mathbb R^{2d}$.   
       
       \bigskip
       We are just left with the calculation of the matrix elements of the metaplectic operator 
       between two different coherent states. We have thus established the following important 
       result \cite{coro5} (here everything is independent of $\hbar$ and we have denoted
       $g_{X}:= \hat T^{(\hbar=1)}(X)g$):
       
        \begin{lemma}
           The matrix elements of $\hat R^1(F)$ on coherent states $g_{X}$ are given by the
           following formula:
           $$\langle g_{Y+\frac{X}{2}}\vert \hat R^1(F)g_{Y-\frac{X}{2}}\rangle =
            $$
           \beq
           \label{met}
           2^d\left(\det V_{t}\right)^{-1/2}
           \exp\left\{(K_{F}-\1)Y\cdot Y + 
           \frac{i}{2} \sigma(X,Y-K_{F}Y-\tilde K_{F}Y)+
           \frac{1}{4}JK_{F}JX\cdot X \right\}
           \eeq
           where $K_{F}:= (\1+F)(\1+F+iJ(\1-F))$ and $\tilde K_{F}$ is the transpose of $K_{F}$.
            \end{lemma}

            \begin{remark}
            If $\det(\1+F)\ne 0$, then
            $$K_{F}\equiv \left(\1+iJ(\1-F)(\1+F)^{-1}\right)^{-1}$$
            In this case $\hat R^1(F)$ has a smooth Weyl symbol given by the following formula:
            (see \cite{coro5} where we have named this formula the ``Mehlig-Wilkinson
            formula'', according to the physics literature \cite{mewi})
            \beq
            \label{mw}
            R(F,X)= e^{i\pi\nu}\left\vert \det(\1+F)\right\vert^{-1/2}\exp\left(
            -iJ(\1-F)(\1+F)^{-1}X \cdot X\right)
            \eeq
            where $\nu$ is the Maslov index that we have computed exactly.
            \end{remark}

           Moreover we have:
           \begin{lemma}\label{unit}
           For any symplectic matrix $F$, consider the matrix $V_{F}$ defined by
           (\ref{VF}). We have that $\vert \det V_{F}\vert \ge 1$, and $\vert \det V_{F} \vert =1$
           if and only if $F$ is unitary.
           \end{lemma}
           
           Proof of Lemma \ref{unit}: Let 
           $$F= \left(\begin{array}{cc}
           A&B\\
           C&D
           \end{array}\right)$$
           be the 4-block decomposition of the $2d\times 2d$ symplectic matrix $F$. We have
           the following diagonalization property of the Hermitian matrix $iJ$:
           $$iJ = U \left(
           \begin{array}{cc}
           -\1&0\\
           0&\1
           \end{array}\right)U^*$$
           where $U$ is the unitary matrix
           $$U = \frac{1}{\sqrt2}\left(
           \begin{array}{cc}
           \1&\1\\
           i\1&-i\1
           \end{array}\right)$$
           Thus we have:
           $$V_{F} = \frac{1}{2}U\left(\left(
           \begin{array}{cc}2&0\\
           0&0
           \end{array}\right)U^*FU + \left(
           \begin{array}{cc}
           0&0\\
           0&2
           \end{array}\right)\right)U^*$$
           and therefore
           $$\det V_{F} = \det \frac{1}{2}\left\{\left(
           \begin{array}{cc}A+D+i(B-C)& A-D-i(B+C)\\
           0&0
           \end{array}\right)+ \left(
           \begin{array}{cc}
           0&0\\
           0&2
           \end{array}\right)\right\}$$
           $$= \det \frac{1}{2}\left(\begin{array}{cc}
           A+D+i(B-C)&A-D-i(B+C)\\
           0&2
           \end{array}\right)= \det \frac{1}{2}(A+D+i(B-C))$$
           We conclude that (recall that $\tilde A$ is the transpose of the matrix $A$):
          $$ \vert \det V_{F}\vert^2 = \det \frac{1}{4}[\tilde A+ \tilde D-i(\tilde B
           -\tilde C)][A+D+i(B-C)]= \det[\1+ L^*L]$$
           with
           $$L= \frac{1}{2}[A-D+i(B+C)]$$
           where we have used the symplecticity of $F$, namely that
           $$\tilde A C-\tilde C A= \tilde D B-\tilde B D=0$$
           $$\tilde A D-\tilde D A= \1$$
           \\
           \sq\\
         
           \noindent
     End of Proof of Theorem 4.2:
     Putting $ X= 2Y=\frac{1}{2\sqrt \hbar}F_{0,t}^{-1}(z_{0,t}-z_{\delta,t})$ in (\ref{met}), we get 
     (\ref{exponent}).\\
     
     Now the estimate (\ref{Delta}) easily follows from the following:
     
     \begin{lemma}
     Let $$\gamma_{F}(X)= \frac{1}{4}X\cdot \Gamma_{F}X$$
     Then for any $X \in \mathbb R^{2d}$ we have:
     $$\Re(\gamma_{F}(X))\le -\frac{\vert X \vert ^2}{2(1+s_{F})}$$
     where $s_{F}\equiv\Vert F\Vert^2$ is the largest value of $F\tilde F$ ($\tilde F$ being the transpose of the matrix $F$).
     \end{lemma}  
     
     Proof:
     let us begin to assume that $\det (\1+F)\ne 0$. Then we have:
     $$K_{F}= (\1+iN)^{-1}\ \mbox{where}\ N = J(\1-F)(\1+F)^{-1}\ \mbox{is real symmetric}$$
     so we can compute
    $$ \Re(K_{F})= (\1+N^2)^{-1}= K_{F}K_{F}^* \ \mbox{and}\ 
    \Im(K_{F})= -N(\1+N^2)^{-1}$$  
    So we get:
    $$\Re\gamma_{F}(X)= \frac{1}{4}\left((\1+JN)K_{F}K_{F}^*(\1-NJ)X \cdot X
    -2\vert X \vert ^2\right)$$
    By definition of $K_{F}$, we have:
    $$(\1+JN)K_{F}= 2 \left( (\1+iJ)F^{-1}+\1-iJ\right)^{-1}:= 2 T_{F}$$
    We have, using that $F$ is symplectic
    $$(T_{F}^*)^{-1}T_{F}^{-1}= 2(\tilde F^{-1}F^{-1}+\1)$$
    Hence we get:
    $$T_{F}T_{F}^*-\frac{\1}{2}= \left (2 (\tilde F^{-1}F^{-1}+\1)\right)^{-1}
    -\frac{\1}{2}= - \frac{\1}{2(\1 + \tilde F F)}$$
    $$T_{F}T_{F}^*X\cdot X-\frac{\vert X \vert^2}{2}= -\frac{1}{2}(\1+\tilde F
    F)^{-1}X\cdot X\le -\frac{1}{2(1+s_{F})}\vert X \vert^2$$
    and the conclusion follows for $\det (\1+F)\ne 0$, hence for every symplectic matrix $F$
    by continuity.\\
    \sq\\

    \bigskip
{\bf Acknowledgements} We thank Jens Bolte for communicating ref. \cite{bolte} before
publication.

     \end{document}